# A bubble-powered micro-rotor: conception, manufacturing, assembly, and characterization

**Jonathan Kao\*, Xiaolin Wang\*, John Warren\*\*, Jie Xu# and Daniel Attinger#**
\* State University of New York at Stony Brook, Stony Brook NY 11794

\*\*Brookhaven National Laboratory, Upton NY 11973

#Columbia University, New York NY 10027

E-mail: da2203@columbia.edu (Daniel Attinger)

**Abstract** A steady fluid flow, called microstreaming, can be generated in the vicinity of a micro-bubble excited by ultrasound. In this article, we use this phenomenon to assemble and power a microfabricated rotor at rotation speeds as high as 625 rpm. The extractible power is estimated to be on the order of a few femtowatts. A first series of experiments with uncontrolled rotor shapes is presented, demonstrating the possibility of this novel actuation scheme. A second series of experiments with 65 μm rotors micromanufactured in SU-8 resin are presented. Variables controlling the rotation speed and rotor stability are investigated, such as the bubble diameter, the acoustic excitation frequency and amplitude, and the rotor geometry. Finally, an outlook is provided on developing this micro-rotor into a MEMS-based motor capable of delivering tunable, infinitesimal rotary power at the microscale.

## 1. Introduction

Since the 1990's, there have been significant research efforts devoted to the development of micro-motors and actuators. These devices are for instance needed for powering microelectronic devices [1-3] and in biomedical applications, where micro-motors [4] are needed for drug delivery systems, surgical tools, probes and as biomechanical actuators for biological cells [5]. Two approaches can be used to achieve these ends. A first approach is to miniaturize an existing macroscopic concept. With photolithography, micro gas turbines [1-3], a miniature Wankel engine [6], and electrostatic micro-motors [4] have been created. These prototypes are remarkable due to their complexity and high power density, on the order of 100MW/m$^3$.

Miniaturization, however, presents inherent difficulties because the ratio of surface forces to volume forces is inversely proportional to the device size. Therefore, friction, wear, and geometric tolerance issues become difficult to handle at the microscale. A second approach takes advantage of the changing ratio of forces inherent to the miniaturization process. This way, designs of greater simplicity can be produced, exemplified by the bubble-jet printing technology, where an explosive bubble acts as a piston [7], or a recently-demonstrated linear propulsion system driven by a bubble oscillating in a channel [8]. It must be stated that these simpler designs sometimes go along with complex thermofluidic phenomena [9], because of the strong coupling of transport phenomena, and the very small time and space scales involved. The novel rotor actuation concept that we present also takes advantage of this second approach, using the micro-flow in the vicinity of a sonicated micro-bubble to attract a rotor and induce a steady rotation.

The working principle of the micro-rotor presented here can be described as follows. When a gas bubble in a liquid is excited by ultrasound, it oscillates according to the Rayleigh-Plesset equation [10]. For a free-floating bubble, the surrounding liquid participates to these oscillations with a symmetrical, radial motion. In the case where the oscillating bubble is attached to a solid wall, as in the concept proposed here, the motion of the surrounding liquid is not radially-symmetrical anymore and the bubble-liquid interface exhibits a coupling of radial and translational motion, through respective expansion (or contraction) and translation perpendicular to the wall. As a consequence, a second-order steady flow, called microstreaming, appears as a donut-shaped vortex in the immediate vicinity of the bubble. Microstreaming was recently investigated theoretically [11] and exploited to control the deformation and rupture of biological vesicles [12]. The low Reynolds number fluid flow characterizing microstreaming can be simulated by superposing potential solutions for Stokes flow, according to an approach briefly sketched in [5, 12], that we developed in details in [13]. The flow streamlines (shown in Figure 1) correspond to a vortical flow structure that is symmetric about the perpendicular axis to the surface where the bubble is attached. The analytical solution sketched in Figure 1 agrees with flow visualization using particles as tracers [13], with a maximum velocity magnitude on the order of 1 mm/s. As described in the experiments below, this microstreaming flow can be exploited to assemble a micro-rotor on the surface of a bubble and control its rotation speed. This process works with two kinds of rotors: rotors with uncontrolled shapes (section 2) and controlled, micromanufactured shapes (section 3 and 4).

## 2. Feasibility of a bubble-powered micro-rotor

This section describes the experimental setup and a first series of experiments that demonstrate the feasibility of actuating a micro-rotor by a sonicated bubble.

*2.1. Experimental setup*

The experimental setup is shown in Figure 2. Experiments take place in a 10 mL glass cuvette with a piezoelectric ceramic transducer (PZT) glued to its bottom. The transducer allows for the generation and control of a standing pressure wave in the fluid contained in the cuvette. In the cuvette, a resistance microheater (Heraeus, Germany) generates isolated micro-bubbles on its surface by gas desorption from the gently heated liquid. The fluid temperature increase $\Delta T$ due to the microheater use can be estimated by equating the enthalpy change of the liquid in the cuvette with the electrical power given to the resistor. This gives $\rho V c_p \Delta T = UI$, with $V = 10$ mL, $U = 5$ V and $I = 0.5$ A. Using the physical properties of water and a heating time of 5 s, an average temperature increase 0.3 K is calculated, which is negligible.

A 15 MHz waveform generator connected to a 500kHz amplifier (Krohn-Hite 7600 M, 200 V) provides an AC voltage $V_{trans}$ to the piezoelectric transducer. A progressive scan camera (Pixelink PL-A741) coupled to a Mitutoyo 70XL microscope zoom is used to observe the motion of the rotor, with temporal and spatial resolutions on the respective order of one µs and one µm and frame rates ranging from 20 to 330 fps. A second function generator is used to pulse a diode to image the oscillations of the bubble driven by the standing pressure wave.

*2.2. Preliminary experimentation with polymer spheres and debris*

In order to visualize the flow, polymer spheres were mixed with water in the cuvette chamber. Using acoustic waves at 180 kHz and bubbles about 20 µm in radius, steady microstreaming flows were observed around the acoustic bubble. The flow pattern was found to correspond to the analytical flow pattern as shown in Figure 1.

Flow visualization experiments were performed with polymer spheres (Eichrom Technologies Inc., Pre_Filter Resin) introduced in water to mimic more precise Particle-Image-Velocimetry (PIV) methods [14]. The 110-150 micrometer particles were crushed in order to obtain smaller particles that would better follow the flow. The general flow pattern was found to agree quantitatively and qualitatively [13] with the simulation in Figure 1. Remarkably, we found that polymer debris [13] such as the slightly asymmetric spherical cap depicted in Figure 3, were attracted to the surface of the bubble upon which they would begin to rotate, as shown in Figure 3. This process was repeated with debris of various shapes and sizes comparable to the bubble, and occurred according to the three following steps: (1) a debris caught in the vortical microstreaming flow moves towards the bubble along one of the streamlines depicted in Figure 1, (2) the debris self-centers on the surface of the bubble, and (3) the polymer would rotate on the bubble's axis of symmetry, most likely do to the symmetrical nature of the microstreaming flows. These three steps describe a self-assembly and actuation process that takes approximately 1 second. The rotor pictured in Figure 3 rotated at a speed of 300 rpm [13]. Also, experiments in [13] showed that the rotor frequency could be controlled by varying the acoustic frequency. This first series of experiments showed the possibility of actuating a micro-rotor by acoustic waves, but did not allow the control and measurement of the rotor geometry.

## 3. Manufacturing of micro-rotors with controlled geometries

A second series of experiments was performed with micro-manufactured rotors. Three types of micro-rotors were designed with diameters of 65 μm and geometries as in Figure 4. The rotor geometries are bio-inspired, mimicking the general shape of the winged seed (samara) of a Stony Brook maple tree. One type is a single layer rotor (A1) while the other two types (A2 and B2) have two layers to better imitate the vein visible on the winged seed. Three batches of micro-rotors were fabricated using optical lithography to pattern SU-8, an epoxy-based photoresist used to fabricate high aspect ratio microstructures for MEMS applications. Due to the delicate optical alignment of the second mask with respect to the first, there is significant misalignment between the two layers in A2-type rotors, while B2-type rotors have only a slight misalignment. A Dektak profilometer measured the thickness of A1-type rotors to be 1.87 μm while B2-type rotors have a thickness of 1.98 μm for the layer in contact with the bubble (Figure 1) and a second layer thickness of 2.20 μm. Thousands of micro-rotors geometry are patterned on separate silicon wafers covered with a 17 nm release layer (XP-SU8, Microchem Corp). The rotors were released by submerging the wafers in an ammonium hydroxide solution and using a water micro-jet. The solution with the floating rotors was then neutralized with sulfuric acid to pH 7. A Brookfield viscometer measures the solution viscosity to be 1.1 $\pm$ 0.1 cP, a value close to the viscosity of water.

## 4. Actuation and characterization of the micro-fabricated micro-rotors

This section describes the actuation process of the micro-rotors and characterizes the parameters affecting rotation speed and stability.

*4.1. Micro-rotors assembly and actuation*

The assembly and actuation of the micro-rotors was done using the procedure detailed in section 2.2. Since thousands of rotors were floating in the cuvette, the assembly could be made in a quick and reproducible manner. The observed direction of the rotation is shown by arrows in Figure 4. Using an appropriate intensity and frequency of ultrasound, a rotor would stay for several minutes on top of the bubble.

*4.2. Characterization*

Our characterization study identified that the rotation speed $n$ depends on the following parameters: the angle $\alpha$ between the rotor plane and the solid surface plane; the bubble radius $a$; the driving ultrasonic frequency $f$; the voltage applied to the piezoelectric transducer $V_{trans}$; and the type of micro-rotor geometry (A1, A2, or B2 as shown in Figure 4). Rotation speed and rotation angle were obtained by optical inspection of frames extracted from the high-speed visualization. The largest rotation frequencies and most stable experiments were made with ultrasound excitation frequencies corresponding to the natural frequency of the cuvette (161.4 kHz) and with 20 micrometer radius bubbles.

*4.2.1. Influence of ultrasonic excitation frequency*

The *frequency* response of micro-rotors was assessed by varying the frequency of the acoustic excitation in the vicinity of the frequency maximizing the rotation speed. Figure 6 depicts the respective frequency response of type B2 micro-rotors. Similar curve shapes were obtained for the other two types of rotors (A1 and B1), although with lower maximum rotation frequencies. Clearly, the rotation speed is maximized when the excitation frequency matches the resonant frequency of the bubble-cuvette system. The frequency response curve is analogous to the frequency response of linear spring-inertia-dashpot system [15], where the response is relatively constant for under-critical excitation frequencies, peaks at the natural frequency, and vanishes for over-critical excitation frequencies. In the system described here, the surface tension is the spring, and the inertia and dashpot correspond to the complex fluid motion and interaction with the propeller. In the case of the micro-rotor, over-critical excitation frequencies would sometimes correspond to cases where the rotor stops turning. Obviously, the rotation speed is very sensitive to the excitation frequency. This high sensitivity is interesting since excitation frequency is a parameter easy to control precisely.

*4.2.2. Influence of the bubble radius*
Our experiments in Figure 5 describe the relation between the rotor rotation speed and the bubble radius $a$ increases, for a constant excitation frequency. This relation can be explained by considering that [10] the bubble radius $a$ and bubble natural frequency $f_b$ are related by the relation $a f_b = 3$ *m/s*. This relation predicts that a bubble with a resonance frequency that matches the cuvette resonant frequency of 161.4 kHz has a radius of 19 μm. It is a reasonable guess to assume that exciting a bubble at its resonance frequency would maximize the associated microstreaming and the rotation speed of the rotor.

Our experiments in Figure 5 show that the micro-rotor rotational speed decreases with increasing bubble radius. While this trend is consistent with the theoretical analysis above that states that $n$ should be maximal for a 19 micrometer bubble, we were not able to obtain more data points because of the difficulty of controlling bubble size in our experimental setup. For instance, no experiments could be made with bubble smaller than the optimal bubble radius. This is due to the fact that the rate of interfacial diffusion of the bubble gas into the water increases with bubble curvature [10], causing smaller bubbles to shrink and disappear within seconds. Therefore, it is reasonable to claim that the relation in Figure 5 between the bubble radius and the rotor rotation speed was obtained with bubbles larger than their optimum size, i.e. with bubbles having a natural frequency lower than the excitation frequency.

*4.2.3. Voltage applied to the piezoelectric transducer*
In Figure 7, the rotation speed is plotted as a function of the voltage applied to the transducer, for a rotor of type A2. The rotation speed increases monotonically with the transducer voltage $V_{trans}$, which can be explained by the fact that the energy delivered to the system increases. A linear regression best fit of the experimental data for A2-geometry gives the following equation $n=6.18V_{trans} - 311$ [rpm]. It must be mentioned that the streaming speed usually grows quadratically with the ultrasonic intensity. This quadratic relation is characteristic of the second-order streaming flow [11, 16]. Given the narrow range of voltages used in Figure 7, it is difficult to determine if the voltage-rotation speed relationship is linear or quadratic. The reason our measurements were performed only for a relatively narrow interval of voltages is that voltages higher than 80V would make the rotor wobble and leave the bubble. It must also be mentioned that more direct ways exist to measure the energy delivered to the system, such as a needle hydrophone, but this device was not available for our measurements.

*4.2.4. Influence of rotor geometry*

High-speed visualization and measurements were made with the three types of rotors to determine what rotor geometry (type A1, A2 and B2 as shown in Figure 4) would work best. The rotor geometry was found to influence the rotation in three ways: the regularity of the rotation, the rotation angle $\alpha$ between the rotor plane and the solid surface plane, and the rotation speed $n$. In most cases, a large value of $\alpha$ would correspond to unstable rotation and detachment of the rotor from the bubble. A1-type rotors were observed behaving erratically, e.g. making a revolution one direction, and then revolving in the reverse direction, in addition to changing its contact point over the surface of the bubble. A2-type rotors typically rotate with $\alpha = 15°$ to $30°$, probably due to the misalignment between the two layers, but never span erratically. Finally, the best results were obtained with B2-type rotors: they rotate with $\alpha = 0°$ and with high regularity. Furthermore, the maximum rotation speeds observed for the A1, A2, and B2 rotors were 33 rpm, 300 rpm, and 625 rpm, respectively. These results, confirm that geometry is a critical parameter for rotor stability and maximizing rotor rotational speeds. Furthermore, these results show that the double layer design, which more accurately mimics the winged seed shown in Figure 4, is more stable and rotates about 20 times faster than the single layer design. Also, our results reveal that misalignment between the two layers should be minimized for stability and efficiency purposes, as exemplified by the better stability and rotation speed of rotors B2 in comparison with rotors A2. To this end, it is evident that optimizing rotor geometry is critical to maximizing rotational speed and extractible power. This is however not a trivial task, because the physics of the problem involves the complex coupling of high-frequency, transient ultrasound and free surface oscillations with a steady, second-order, highly viscous flow, as well as the interaction of this complex flow with a complex solid shape that is free to move.

*4.3. Power output*

The power output of these rotors can be estimated as follows. At low Reynolds numbers, the power given by the fluid to the rotor is approximately equal to the viscous dissipation that would be produced by the rotation of a similar rotor, at a similar speed, in a quiescent fluid. Von Karman solved the laminar flow created by a rotating disk, and the corresponding torque on one side of the disk is given in [17] as $M = 0.62 \frac{\pi}{2} \rho a^4 \sqrt{\upsilon \omega^3}$, where $\rho$ is the density and $\nu$ is the kinematic viscosity of the surrounding fluid. The rotary power can therefore be estimated as $P = 2M\omega$. Using the physical properties of water, the radius of the manufactured rotor, and a rotational speed of $\omega$ = 65/s (625 rpm), we obtain a power output to be on the order of $P$ = 74 femtowatt. This is obviously a very low power, however it is obtained by a very simple microdevice and can be precisely tuned by e.g. the excitation frequency.

**5. Outlook**

A tentative design to extract rotary power from the rotor presented in this article is proposed in Figure 8. A sub-micrometer diameter shaft can be attached perpendicularly to the rotor to transmit the mechanical power through the bubble and the solid wall. Multiwall carbon nanotubes can be used to build the shaft, because they have a sub-micrometer diameter and extremely low relative friction. Therefore, an outer tube will be glued to the wall where the bubble sits, while the inner tube will remain attached to the rotor. Alternatively, biofibers can also be used. This system may also be scaled down, provided that the bubble is sufficiently stable to degassing [10], an issue which can be addressed with surfactants and the use of different gases and liquids. Furthermore, because only one dimension is required to control the frequency of a standing acoustic wave, concentric geometries can be used to minimize the total volume of the micromotor as shown in Figure 8a. It is therefore possible to fit several motors in series in a space as thin as human hair, as pictured in Figure 8b. Finally, remote activation (through acoustic wave) allows design and activation of micro-motors in parallel, as well as selectivity between bubbles of different sizes. By conveniently controlling the excitation amplitude and frequency, the proposed micro-motor solutions will be capable of delivering infinitesimal torque at the microscale.

## 6. Conclusions

The feasibility of an acoustic, bubble-powered micro-rotor is demonstrated. A steady vortical microstreaming flow is used to self-assemble and actuate a micro-rotor on the surface of the bubble. Measurements with a high-speed camera have determined how the rotation speed and stability is affected by the bubble radius, the excitation frequency and voltage, and the rotor geometry. Our measurements show that the frequency response of the rotor to the excitation frequency is analogous to the frequency response of a linear spring-inertia-dashpot system. High sensitivity to the excitation frequency provides a convenient way to control the rotation speed. Finally, designs are presented that will allow the use of this micro-rotor concept to apply controllable and infinitesimal torque at the microscale.

**Acknowledgements**

This research was supported by the NSF CAREER grant 0449269**.** This research was carried out in part at the Center for Functional Nanomaterials, Brookhaven National Laboratory, which is supported by the U.S. Department of Energy, Division of Materials Sciences and Division of Chemical Sciences, under Contract No. DE-AC02-98CH10886. Jonathan Kao was supported by the Simon's fellowship program at Stony Brook University. Daniel Attinger thanks Howard Stone, from Harvard University, for useful discussions on Stokes flows.

**Figures**

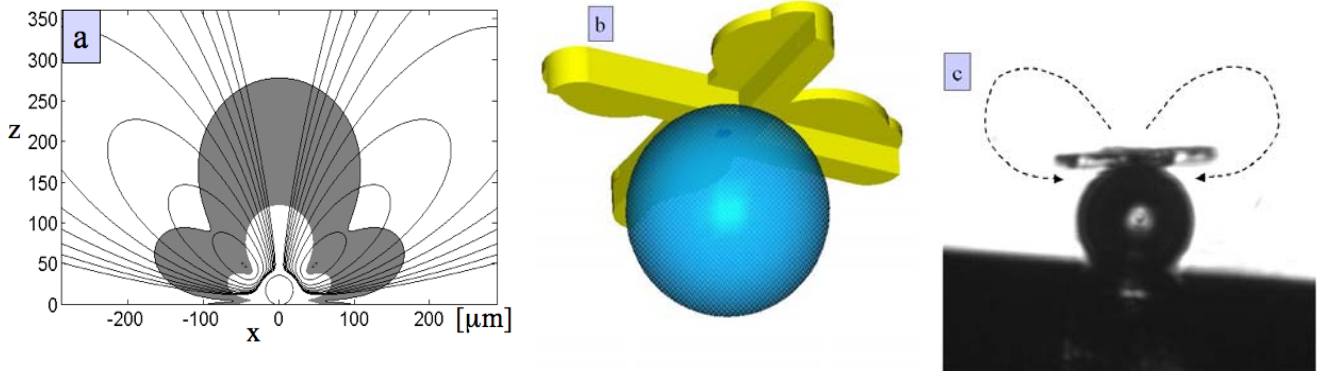

Figure 1: (a) Streamlines corresponding the analytical solution of the microstreaming flow around a bubble with 18 mm radius [13]. The gray zone corresponds to velocities between 1mm/s (inner boundary of gray zone) and 0.1 mm/s (outer boundary of gray zone). (b) 3D model of the micro-rotor and the bubble (c) Photograph of the actual bubble-actuated micro-rotor. The microstreaming vortical flow streamlines are outlined by the dotted lines superimposed on the image. The rotor diameter is 65 μm. A movie of the rotor in action is available in [18].

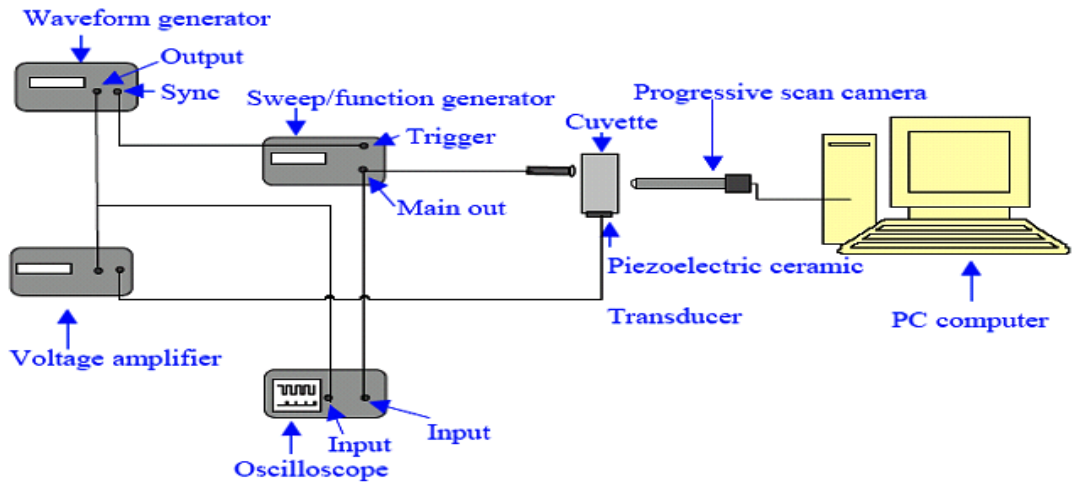

Figure 2: Experimental setup. The function and waveform generator are used to drive the illumination and piezoelectric excitation. The cuvette contains the micro-bubbles and the micro-rotors.

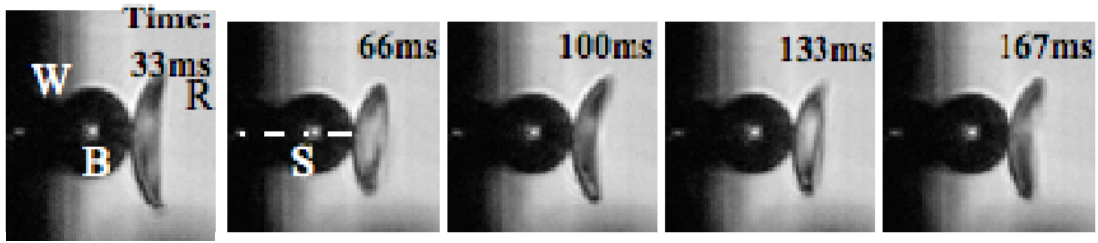

Figure 3: A slightly asymmetrical polymer rotor rotates at 300 rpm. The letters W, B, R, and S correspondrespectively to the wall on which the bubble sits, the microbubble, the rotor, and the proposed shaft detailed in Section 5.

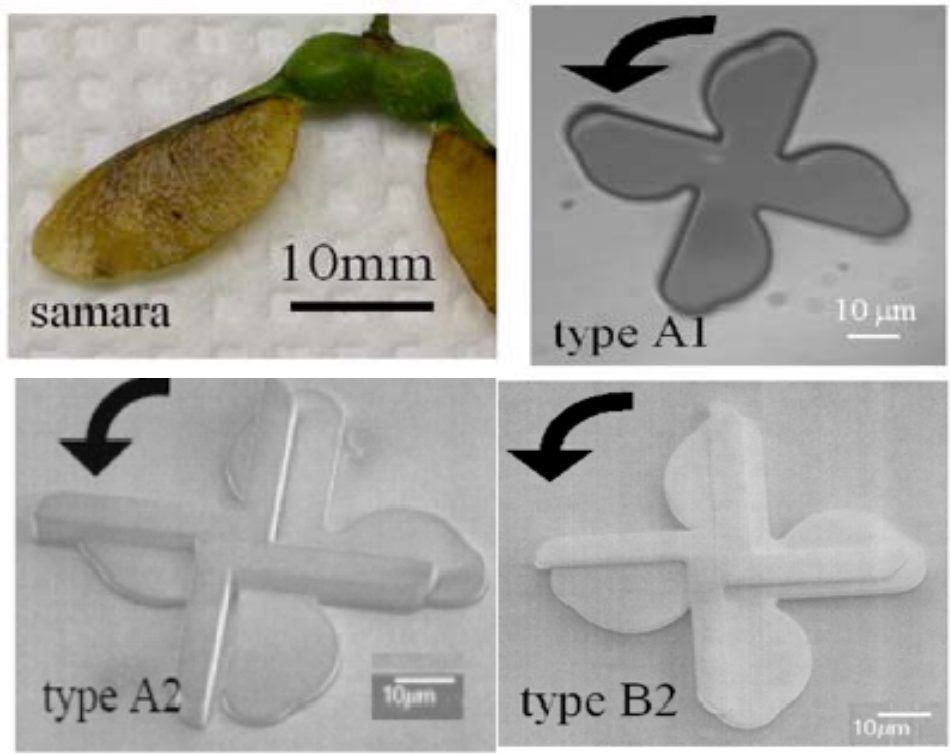

Figure 4: The geometry of the maple tree fruit or samara (top left), which inspired the rotor design. A1-type rotors are single layered. A2 and B2 rotors are double-layered; A2 has significant misalignment between the two layers, while B2 has less misalignment.

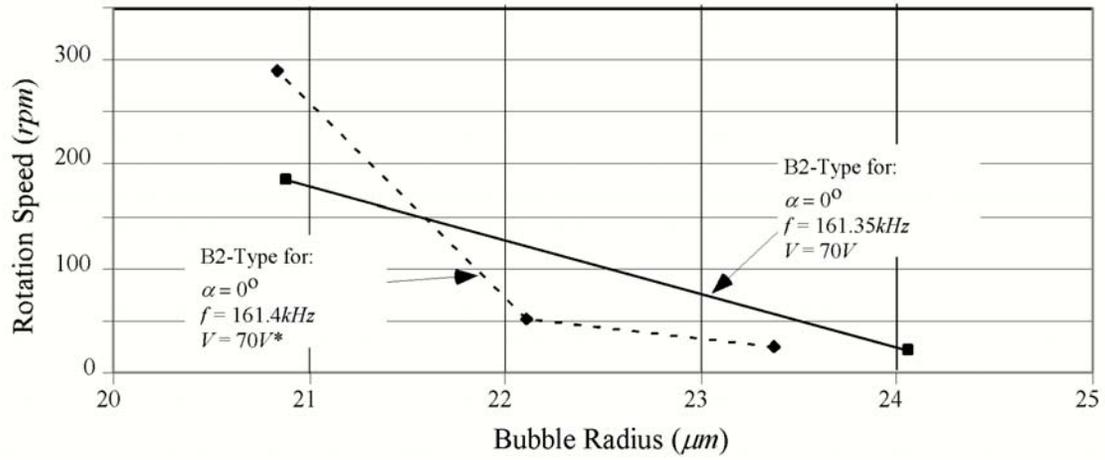

Figure 5: Measurements describing the effect of bubble radius on rotation speed. Results show that rotation speed decreases as the bubble radius increases.

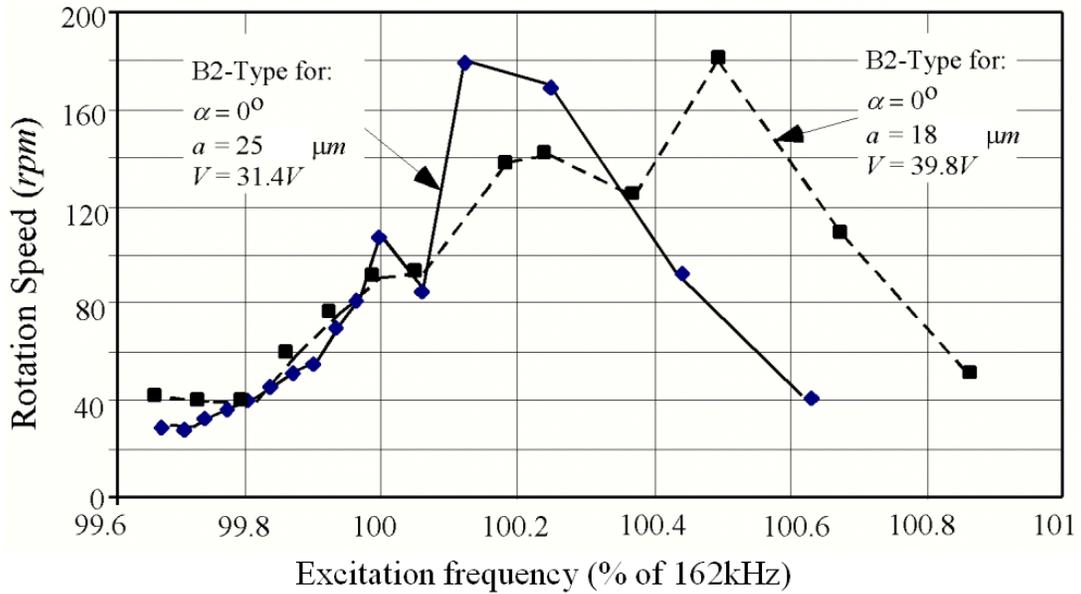

Figure 6: Influence of the excitation frequency on the rotational speed with B2-type rotors. The speed is measured from high-speed visualizations.

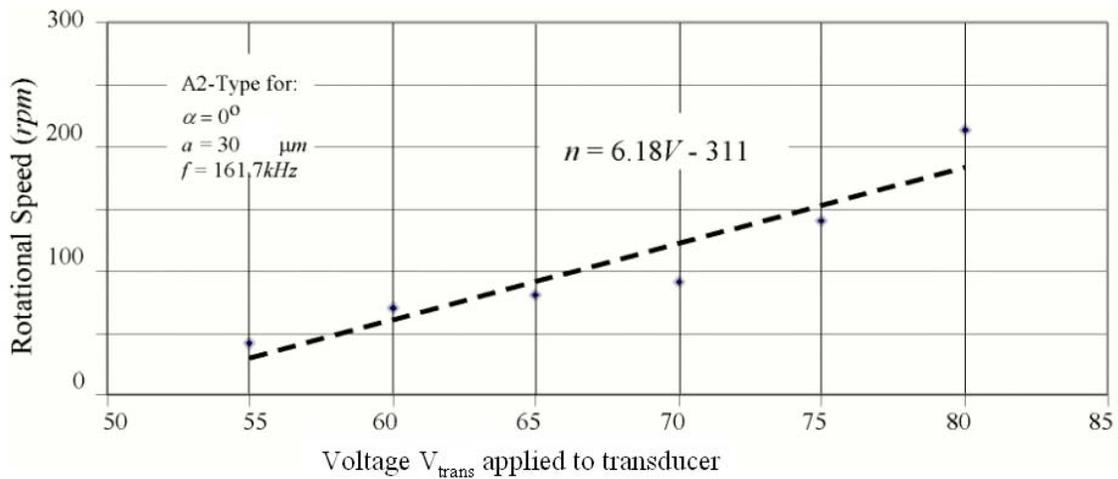

Figure 7: Rotational speed of a micro-rotor as a function of the voltage applied to the piezoelectric transducer.

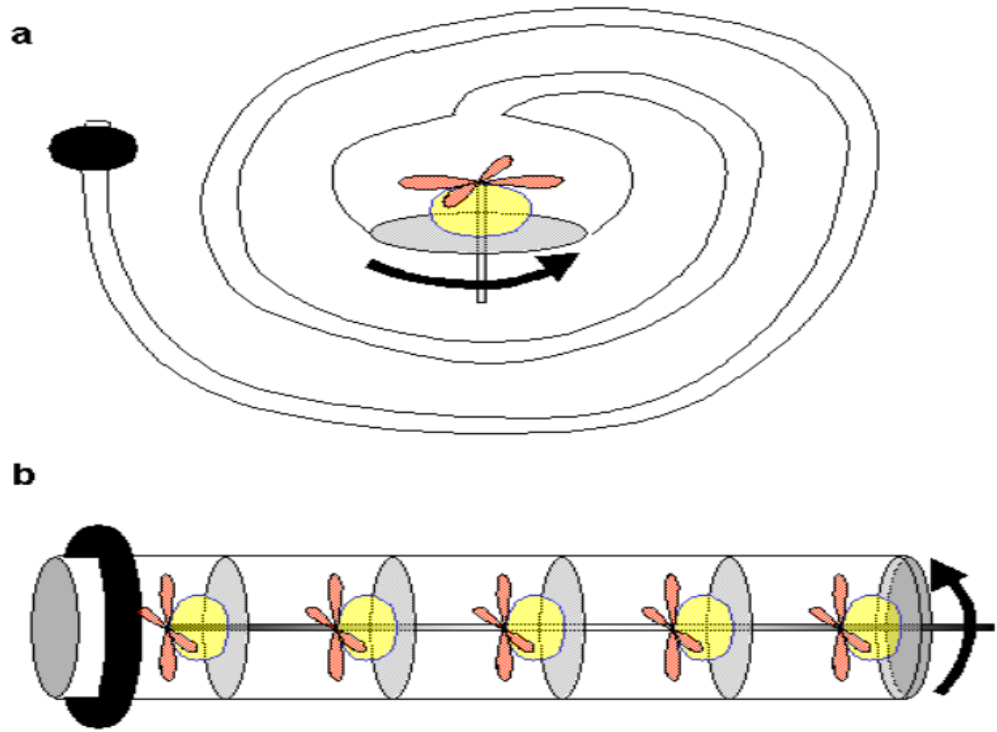

Figure 8: Tentative packaging of the micro-motor. The thick ring at the top is a piezoelectric transducer. In (a), a concentric geometry reduces the volume. In (b), five motors in series drive a unique shaft.